\documentclass{iauc}
\usepackage{graphicx}
\pubyear{2005}
\volume{199}
\pagerange{1--}
\jname{Probing Galaxies through Quasar Absorption Lines}
\editors{P. R. Williams, C. Shu, and B. M\'{e}nard, eds.}

\title[DLA Surveys] 
{Damped Lyman $\alpha$ Surveys and Statistics:\\
 A Review}

\author[S. Rao]   
{Sandhya M. Rao}

\affiliation{Dept. of Physics and Astronomy, University of Pittsburgh, 
Pittsburgh, PA 15260, USA\break 
email: rao@everest.phyast.pitt.edu}

\begin{document}

\newcommand{\Wmi}{$W_0^{\lambda2796}$}
\newcommand{\Wf}{$W_0^{\lambda2600}$}
\newcommand{\Wmii}{$W_0^{\lambda2803}$}
\newcommand{\Wmiii}{$W_0^{\lambda2852}$}

\maketitle

\begin{abstract}

The discovery of the first damped Lyman alpha (DLA) system in the
early 1970s followed by the recognition that DLAs arise in intervening
galaxies opened up a new field of galaxy evolution research. These
highest HI column density absorption-line systems trace the bulk of
the observed neutral gas in the universe, and therefore, have been
used as powerful probes of galaxy formation and evolution back to the
redshifts of  the most distant quasars. The history and progress of
DLA research over the past several decades is reviewed here. Larger
datasets and deeper surveys, particularly over the last couple of
years, have improved our knowledge of the neutral gas content and
distribution in the universe at all observable redshifts, including
the present epoch. New results on the statistics of DLAs at $z<1.65$
from our HST-UV surveys are presented and discussed in  the context of
recent results at $z=0$ and at high redshift. We find that
$\Omega_{DLA}(z>0)$ remains roughly constant to within the
uncertainties; the $z=0$ value of the neutral gas mass density,
$\Omega_{g}$,  is a factor of $\approx 2$ less than $\Omega_{DLA}$.
The DLA incidence, $n(z)$, undergoes rapid evolution between redshifts
5 and 2, but is consistent with the no-evolution curve in the current
concordance cosmology for $z\lesssim 2$. We also show that if the
local Schmidt law relating surface density of gas and star formation
rate (SFR) is valid at the DLA redshifts, then  the DLA SFR density is
too low for them to provide a significant contribution to the cosmic
star formation history (SFH) at $z\gtrsim 1$.  This implies that the
DLAs are unlikely to be the same population as the
star forming galaxies (i.e., the Lyman break and sub-millimeter galaxies)
 that dominate the  SFH of the high redshift universe. We suggest
that this discrepancy and the DLA ``missing metals'' problem could 
be the result of missing very high column density gas
due to its very small absorption cross section.

\keywords{(galaxies:) surveys, quasars: absorption lines, galaxies:
statistics}

\end{abstract}

\firstsection 
\section{Introduction}

The first published spectrum of a damped Lyman $\alpha$ (DLA) system,
the $z=2.309$ absorber towards PHL 957   by Lowrance et al. (1971,
1972), was  quickly followed by Beaver et al.'s (1972) independent
observation  of this object and their interpretation that the strong
absorption line might be due to radiation damping from very high
column density HI. They showed that the line profile was consistent
with a column density of $N(HI)=2\times 10^{21}$ atoms cm$^{-2}$, and
noted its similarity to Ly$\alpha$  absorption features along lines of
sight to Galactic stars (Savage  \& Jenkins 1972). Serendipitous
discoveries of DLAs towards the  quasars Q1331+170, PKS 1157+014, and
PKS 0528$-$250 were reported shortly thereafter (Carswell et al. 1975;
Wright et al. 1979;  Smith et al. 1979). The discovery of DLAs in the
early 1970s in combination with  the realization  that  Ly$\alpha$
forest and narrow metal absorption lines arise in intervening material
(Wagoner 1967; Bahcall \& Spitzer 1969; Lynds 1971; Weymann et
al. 1979;  Sargent et al. 1980), led to a whole new approach to the
study of galaxy evolution. By tracking absorption line properties with
redshift, it was now becoming possible to  explore the evolution of
gaseous structures in the universe.  DLAs are the highest column
density neutral hydrogen absorption lines seen in quasar spectra
($N(HI)\ge 2\times 10^{20}$ cm$^{-2}$), and consequently,  have been
described as harboring the bulk of the neutral gas in the
universe. Thus, barring any selection  effects, tracking  their
properties as a function of redshift  directly  traces the cosmic
evolution of neutral gas in the universe, and since  cold neutral gas
is a requirement  for subsequent star formation, DLAs  were soon
recognized as important probes of galaxy formation  and evolution
(Wolfe, Turnshek, Smith, \& Cohen 1986). As we describe below, Art Wolfe  and
collaborators were the first to exploit the  potential of DLAs as
cosmological probes, and in the early 1980s, began the first
systematic search for DLAs.

Prior to the deep imaging surveys of the last decade, DLAs were the
only  tracers of galaxies at high-redshift. A DLA system was also considered
to be an unbiased probe since its detection was independent of the
luminosity of the absorbing galaxy, and therefore, DLA-selected
samples were thought to be more representative of the  galaxy population. The
detection of a DLA system only required the presence of a background quasar
of sufficient brightness so that its spectrum could be
obtained. Moreover,  the strong line with damping wings characteristic
of a DLA system could be  detected in low-resolution spectra with
signal-to-noise ratios not much greater than a few. In other words,
they were ideally suited for searches through spectroscopic surveys of
quasars.

\section{DLA Surveys: A brief history}

Since the redshifted Ly$\alpha$ line can be observed with groundbased
telescopes only for redshifts $z>1.65$, initial DLA studies   were
restricted to this regime. Redshifted 21 cm observations  of radio
loud quasars have also been used to detect large columns of  neutral
gas in absorption (see Briggs 1999 for an excellent review). Until the
advent of UV telescopes, 21 cm absorbers along the line of sight to
radio loud quasars were the only known high $N(HI)$ systems below
redshift 1.65. Discovery  of the first redshifted 21 cm absorption
line was reported by Brown \& Roberts  (1973) towards 3C 286 at
$z=0.692$,  not long after the first optically  discovered DLA, in a
blind survey towards 100 (radio loud) quasars.  The DLA towards 3C 196
was also discovered in this survey.  Several more were detected
towards quasars that had optically-identified  low-ionization
absorption lines of MgII (Briggs \& Wolfe 1983;  see Briggs 1988 for a
summary).  By the early 1980s the pace of DLA discoveries was picking
up, but it was clear that surveys were required for DLAs to become
useful as statistical probes of neutral gas in the universe. In this
section, we describe the progress of DLA studies beginning with
optical surveys.  It was also obvious that since our end point for
galaxy evolution was all around us at $z=0$, the neutral gas
distribution at the  present epoch could be used to anchor the
evolutionary properties of  galaxies observed in the distant
universe. Rao \& Briggs (1993) made the first attempt to describe the
neutral gas distribution at $z=0$ in a framework for comparison with
high redshift DLA studies.  In \S2.2, we describe this and subsequent
$z=0$ surveys for ``DLA'' gas.  Finally, in \S2.3, we present  results
from surveys for $z<1.65$ DLAs, the most recently explored  redshift
regime.

\subsection{Optical surveys for DLAs}

The Lick DLA Survey was the first systematic search for DLA candidates
at redshift $z\sim 2$ (Wolfe et al.  1986). Wolfe
et al. characterized the survey as a search for high redshift HI
disks, motivated by the fact that DLAs had HI column densities
comparable to those found in the Milky Way and in nearby spiral
galaxies. The similarities to HI disks extended to their 21 cm
properties, i.e., low spin temperatures, high 21 cm optical depths,
and low velocity dispersions. The DLAs also showed low-ionization
metal-line transitions such as MgII, FeII, CII, SiII, OI, etc.,
characteristic  of cold clouds in the Milky Way's ISM. The survey
included 68 quasars whose  spectra were obtained at Lick Observatory
at 10 \AA\ resolution. This resulted in 47 absorption lines tagged as
DLA candidates in a total redshift path  of $\Delta z = 55$. While
this was an unexpectedly high incidence of DLAs given the then known
incidence of Ly$\alpha$ forest systems (Sargent et al.  1980),
follow-up higher resolution spectroscopy (1-2 \AA) confirmed 15
bonafide DLAs (Turnshek et al. 1989; Wolfe et al. 1993). This led to
the first  determination of the  incidence of DLA systems, i.e., the
number of DLAs per unit redshift. They found $dn/dz=0.29\pm 0.07$ at
$\left< z \right>=2.5$.

This was followed by a larger survey of 100 additional quasars
by  Lanzetta et al. (1991). Including results from the Lick survey,
they  found 38 confirmed DLAs in 156 spectra over the redshift range
$1.6\!<\!z\!<\!4.1$ and a redshift path of $\Delta z\!=\!155$. With this larger
DLA sample, they were able to determine for the first time the key
statistical properties that are now routinely used to characterize the
evolution of neutral gas, namely, the redshift number
density $dn(z)/dz$ [or $n(z)$],  the cosmological neutral gas mass
density $\Omega_g(z)$,\footnote{$\Omega_g=1.3\Omega_{HI}$, where the
factor 1.3 corrects for a neutral gas composition of 75\% H and 25\%
He by mass.}  and the column density distribution $f(N)$ for the
DLAs. Any theory of galaxy evolution could now be constrained by the
observed properties of the  neutral gas distribution in the
universe. Lanzetta et al. found that  the redshift number density
could be expressed as a power law  $n(z)=n_0(1+z)^\gamma$ with
$\gamma=0.3\pm 1.4$, which implied that  the incidence of DLAs did not
evolve with redshift in the non-$\Lambda$ cosmology that was assumed
at the time. They also found that $\Omega_g(z=2.5)\approx 1 \times
10^{-3}$ was comparable to the mass density of stars in present-day
spiral galaxies, and that it did not evolve over the redshifts probed
by the survey. Their $f(N)$ distribution was a power law, $f(N)\propto
N^{-\beta}$, with $\beta=1.67$, and showed an  excess of systems
above the DLA threshold in comparison to the  extrapolated power-law
distribution of  lower $N(HI)$ systems.

Shortly thereafter, Wolfe et al. (1995) described results from
spectroscopy of 228 quasars drawn from the Large Bright Quasar
Survey.  Their statistical sample now included 58 confirmed DLAs at
$z>1.65$ in a redshift path length of over 300. In contrast to the
Lanzetta et al. (1991)  result, they found that $\Omega_g$ decreased
with decreasing redshift, and interpreted this evolution as due to
depletion from star  formation. In addition, they modelled the
evolution in the $f(N)$ distribution  as being due to star formation
in randomly oriented disks and found agreement with  the observations
if the SFRs followed the standard Schmidt-Kennicutt
law seen in nearby galaxies.

The next advancement in DLA studies from the ground occurred when
surveys were extended to higher redshifts. As high redshift quasars
became more numerous, so did high redshift DLAs; Storrie-Lombardi et
al. (1996a,b), Storrie-Lombardi \& Wolfe (2000), and P\'eroux et
al. (2001, 2003) extended  this work beyond $z=4$. Storrie-Lombardi \&
Wolfe compiled a list of 81 DLAs in the redshift interval
$1.65\!<\!z\!<\!4.69$; they found that $\Omega_g$  was consistent with 
being constant for $2\!<\!z\!<\!4$ and marginally  consistent
with a decline at $z\!>\!4$. P\'eroux et al. discovered an additional 26
new DLAs, 15 at $z\!>\!3.5$, in their survey of 66 $z\gtrsim 4$
quasars. The highest redshift DLA now stood at $z=4.46$. P\'eroux et
al. also concluded that only 55\% of  the neutral gas mass at  $z>3.5$
was in DLAs and that the remaining was contained in Lyman limit and 
sub-DLA systems with HI column densities $10^{19}<N(HI)<2\times 10^{20}$
cm$^{-2}$. Their result that the total neutral gas mass is conserved
over redshifts $2\!<\!z\!<\!5$ led them to suggest that the $z\approx
3.5$ turnover in $\Omega_g$  was a signature of the epoch of formation
and initial  collapse of DLAs.

With the advent of the Sloan Digital Sky Survey (SDSS) and its
database of tens of  thousands of quasar spectra, the next logical
step was to mine it for DLAs.  Prochaska \& Herbert-Fort (2004) used
the SDSS Data Release 1 to find 71 DLA systems in 1252 quasar spectra
at a mean redshift of $z\approx2.5$. Contrary to  the result of
P\'eroux et al., they found that Lyman limit systems   contribute less
than 15\% to $\Omega_g$ at $z>3.5$. This conclusion was based on the
detection of 6 new DLAs at $z>3.5$, half of which have $N(HI)>1\times
10^{21}$ cm$^{-2}$. However, as noted by the authors themselves, the
SDSS is most sensitive to DLA redshifts between 2.1 and 3, and the
statistical uncertainties beyond $z=3.5$ are still too high to make a
definitive statement about the relative contributions of DLAs and
sub-DLAs to the highest redshift values of $\Omega_g$. Prochaska \&
Herbert-Fort also found that $n(z)$ is consistent with previous
estimates, and that $\Omega_g$ decreases with decreasing redshift from
$z\sim 4$ to $z\sim 2$ with a possible decline beyond $z\sim 4$.
However, all of the data points in their figure 5 are within
$\sim1\sigma$ of each other, implying that $\Omega_g$ is constant
within this redshift range. As we show in  \S2.3, the low redshift
data are consistent with this interpretation.  Now, with their larger
SDSS DR3 sample in which they detect over 700 DLAs beyond $z=2.1$,
Prochaska et al. (2005) report evidence for a statistically
significant  decline in $\Omega_g$ from $z\sim 4$ to $z\sim 2$.  The
reader is referred to J. Prochaska's contribution in these proceedings
for  details on their new results.

Fall \& Pei (1989) first suggested that dust in high column density
absorbers would influence the statistics of DLAs. The concern was that
optically selected quasar samples would miss DLAs due to the reddening
and dimming caused by foreground DLA host galaxies (see also Pei et
al. 1999).  To better evaluate this effect, Ellison et al. (2001)
conducted a survey for DLAs towards radio-selected quasars and
obtained spectra of all quasars regardless of their optical
magnitude. They found that  at the mean redshift of their sample,
$\left<z\right>\!=\!2.4$, both $n_{DLA}$ and $\Omega_{DLA}$ were only
marginally higher than  those measured using optically selected
samples.  Based on the uncertainties in their measurements, which were
dominated by sample size, they concluded that $\Omega_{DLA}$ could be
no more than a factor of two higher than that derived from optical
surveys, and that there was no hidden population of DLAs that was
being missed in optical surveys.   The effects of dust and other
biases that might affect DLA statistics will be discussed further in
\S4. See Turnshek et al. (2005, these proceedings) for more on 
DLA selection effects.

\subsection{The HI distribution at $z=0$}

The first attempt to describe the HI distribution at the present epoch
in a  form suitable for comparison with DLAs was made by Rao \& Briggs
(1993) and Rao (1994) who used HI 21 cm emission observations of  a
complete sample of nearby spiral galaxies to estimate $f(N)$. The 27
galaxies used in the study  included all galaxies with optical
diameters greater than 7' that were in the declination range
observable at Arecibo. They also excluded galaxies in the local
group. This Arecibo survey (Briggs et al. 1980) was initiated to
explore the distribution of HI in the outskirts of local galaxies with
the primary aim of determining the extent of local galaxy halos. The
results were then used to interpret quasar absorption line
properties. For example, Briggs et al. argued that the lack of
detectable HI beyond 2-3 Holmberg radii meant that local galaxies
could not reproduce  the observed high-redshift MgII absorption line
statistics. A decade later, we used these  same data on the radial HI
distribution of local galaxies to calculate $f(N,z\!=\!0)$ as
described in Rao \& Briggs (1993). In brief, if
$<\!\!A(M,N)\!\!>\Delta N$ is the mean cross-sectional area subtended
by gas with HI column density between $N$ and $N+\Delta N$ in a galaxy
of magnitude $M$, then
$$ f(N) = \frac{c}{H_0} \frac{\sum \Phi(M)\left<A(M,N)\right> \Delta N
\Delta M}{\Delta N}.$$ Here, $\Phi (M)$ is the luminosity function of
gas-rich galaxies and is used to normalize $f(N)$.  Since the optical
luminosity of each galaxy in the sample was known and its radial
$N(HI)$ distribution was measured, $\left<\!A(M,N)\!\right> \Delta N$, 
the area averaged
over all inclinations, could be determined. A comparison with the
$\left<z\right>=2.5$ results of Lanzetta et al. (1991)   showed that
$f(N,z\!=\!0)$ was systematically lower than $f(N,z\!=\!2.5)$, and
$n(z\!=\!0)$, which is the integral of $f(N)$ over all DLA column
densities, was a factor of $\approx$5 lower  than $n(z\!=\!2.5)$. The
cosmological mass density of HI, $\Omega_{HI}$, which is proportional
to $\int Nf(N)dN$, was in agreement with other estimates of the
$z\!=\!0$ mass density of  neutral gas (e.g., Fall \& Pei 1993), and
was a factor of 3 less than the $z=2.5$ value.  Thus, we had found
clear evidence for evolution of neutral gas properties to the present
epoch.

However, the local galaxy luminosity function as determined prior to
1993 did not include gas-rich galaxies that are now known to occupy
the low-luminosity tail of the galaxy luminosity function. The overall
normalization  might have been underestimated as well. This offset
manifests itself in the values derived for $f(N)$, $n(z)$, and
$\Omega_{HI}$.  Additionally,  the small sample and the large Arecibo
beam size meant that improvements could be made (and were necessary)
over this initial study. Moreover, the question of whether optically
luminous galaxies contained all the HI was still being debated, and
the possibility that HI  clouds with  no optical counterpart could
contribute to the HI mass and cross section in the local universe was
not completely discounted. To investigate this and to determine $f(N)$
more accurately, many blind HI surveys were launched in the 
1990s. The Arecibo HI Strip Survey (AHISS; Sorar 1994; Zwaan et
al. 1997), the Arecibo Slice survey (Spitzak \& Schneider 1998), the
Arecibo Dual Beam Survey (Rosenberg \& Schneider 2000), the Parkes all
sky southern survey, HIPASS, (Zwaan et al. 2003, 2005; Ryan-Weber et
al. 2003), and most recently, the Westerbork Synthesis Radio Telescope
(WSRT) survey discussed by
M. Zwaan in these proceedings have dramatically improved our knowledge
of the HI distribution at the present epoch.  Apart from determining
$f(N)$, $n(z)$, and $\Omega_{HI}$ very accurately, these surveys
showed convincingly that all of the neutral gas at the present epoch
was accounted for by galaxies included in the optical luminosity
function.  Minchin  et al. (2003), with HIDEEP -- a 20 times more
sensitive survey than HIPASS in a smaller region of sky -- also found
that all their HI detections had optical counterparts. This survey's
limits, $N(HI)\approx 4\times 10^{18}$ cm$^{-2}$ and $M(HI)=4\times
10^5$ M$_\odot$, were the highest sensitivity limits achieved thus far.
Thus, the consensus now is that there are no lurking HI   giants or
dwarfs in the local universe. While the HI mass density and
cross-section are dominated by the high surface brightness, HI rich
galaxies,  Minchin et al. (2004) showed that low surface brightness
galaxies (LSBs) contribute 30\% to the HI mass density and 40\% to the
HI cross-section at the present epoch.  This is very different from
what was assumed in the initial study of Rao \& Briggs (1993), and as
Zwaan (this proceedings) also asserts, is completely consistent with
the high fraction of LSBs identified as DLA host galaxies at low
redshift (e.g.,  Rao et al. 2003).

\subsection{The UV regime: $z<1.65$}

Since the Ly$\alpha$ line falls in the UV for redshifts $z<1.65$, this
most recent 70\% of the age of the universe remained hidden until IUE
and HST. The need for DLA surveys in the UV is illustrated in Figure 1
where $\Omega_{DLA}$ is plotted as a function of time. (We now refer
to $\Omega_g$ as $\Omega_{DLA}$ to specify the cosmological neutral
gas mass density measured using DLAs. This change in notation is
important because, as we discuss in \S4, there is reason to believe
that DLAs do not trace all the neutral gas in the universe.) As
discussed above, both the optical, high-redshift and 21 cm, $z=0$
regimes have been serviced by all-sky surveys, namely, the SDSS and
HIPASS. While an all-sky UV spectroscopic survey does not exist, and
is unlikely, far-fetched, and wishful thinking on our part, we have,
nevertheless, been making incremental progress towards determining DLA
statistics in this regime.  The first concerted effort to obtain a
large sample of UV spectra of quasars was made by the IUE satellite.
Kinney et al. (1991) and Lanzetta, Turnshek, \& Sandoval (1993)
published a total of 260 IUE quasar and AGN spectra, which were
subsequently used by  Lanzetta, Wolfe, \& Turnshek (1995) to estimate
the incidence of DLAs in the UV.  The IUE spectra proved to be
unreliable, and follow-up spectra of the DLA candidates with HST
confirmed only one of the four DLAs that were used to determine DLA
statistics (Turnshek \& Rao 2002). Two were shown to not be DLAs,
while the fourth was not observed with HST. Additionally, one system
which was later shown to be a DLA did not satisfy the IUE sample's DLA
criterion.

\begin{figure}
\centerline{\scalebox{0.5}{\includegraphics{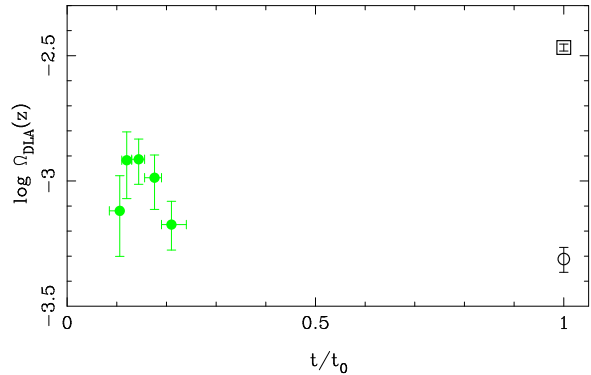}}}
\caption{The log of $\Omega_{DLA}$ as a function of time (for the
``737'' cosmology: $h=0.7$, $\Omega_M=0.3$, $\Omega_\Lambda = 0.7$)
with $t_0$  representing $z=0$. The open circle is $\Omega_g$ at $z=0$
from  Zwaan et al. (2005) and the open square is the mass density of
luminous matter at the present epoch from Panter et al. (2004). The
high redshift (early times) data points are from Prochaska \&
Herbert-Fort (2004). The large gap in time accessible only by UV
surveys for Ly$\alpha$ illustrates the  importance of the $z<1.65$
redshift regime for DLA evolution studies. }
\end{figure}

DLAs at low redshift were turning out be very rare objects, and sample
sizes had to be increased several fold to be able to get reasonably
good statistics. Traditional blind surveys were not yielding useful
results. In fact, the HST QSO absorption line Key Project found only
one DLA  in a redshift path of $\Delta z=49$ (Jannuzi et al. 1998).
In Rao (1994) and Rao, Turnshek, \& Briggs (1995), we used IUE and HST
archival spectra to search for DLAs in previously-known,
optically-identified MgII systems.  This was similar to the selection
procedure used by Briggs \& Wolfe (1983)  to find 21 cm absorbers
towards radio-loud quasars. Since low ionization metal lines are
always seen along neutral gas sightlines in the Milky Way and in
high-redshift DLAs, they act as tracers of high column density
HI. Moreover, the MgII doublet, $\lambda\lambda 2796, 2803$, is
accessible for $z>0.11$ in optical quasar spectra. Beginning in the
late 1980s, several groups  conducted MgII absorption line surveys and
discovered many more MgII systems than were available for the initial
21 cm searches of Briggs \& Wolfe (Lanzetta, Turnshek, \& Wolfe 1987;
Tytler et al. 1987; Sargent et al. 1988, 1989; Caulet
1989; Steidel \& Sargent 1992, SS92).  The SS92 survey, which was the
largest,    included 107 systems and resulted in the determination of
the incidence of MgII systems as a function of redshift and MgII
$\lambda2796$ rest equivalent width. Thus, if a MgII sample was
selected appropriately and the fraction of DLAs in this sample could
be determined, then the  MgII statistics could be used to calculate
DLA statistics. We found 4 candidate DLAs in available IUE and HST
archival data, and were able to place an  upper limit on the values of
$n(z)$ and $\Omega_{DLA}$ at $z=0.8$ by using this method.

In order to increase the low-redshift DLA sample, we obtained UV
spectra of a sample of 60 additional MgII systems with HST-FOS in
Cycle 6. The sample of MgII lines for our initial archival and Cycle 6
surveys was culled from the literature (Rao \& Turnshek 2000, RT00)
and had a rest equivalent width limit of  $W_0^{\lambda2796} > 0.3$
\AA. The total sample of MgII systems with UV Ly$\alpha$ information
now included 82 systems of which 12 were DLAs.  With these results, we
showed that $\Omega_{DLA}$ was high at redshifts $0.1\!<\!z\!<\!1.65$,
consistent with a  constant value for the neutral gas mass density
from $z\approx 4$ to $z\approx 0.5$. This was primarily due to the
high fraction of $N(HI)>10^{21}$ cm$^{-2}$ systems found in the
survey. Our results also suggested that systems with both MgII
$\lambda 2796$ rest equivalent width, $W_0^{\lambda2796} > 0.5$ \AA\
and FeII $\lambda 2600$ rest equivalent width, $W_0^{\lambda2600} >
0.5$ \AA\ had a 50\% chance of being a DLA. Based on this result, we
conducted a similar survey of 54 MgII systems in 37 quasars with
HST-STIS in  Cycle 9. Most of these satisfied the strong
MgII-FeII criterion for DLAs. Twenty seven had useful UV spectra and
four of these were DLAs. The DLA towards Q1629+120 was discovered in
this survey and was reported in Rao et al. (2003). The others are
reported in Rao, Turnshek, \& Nestor (2005a, RTN05).  The strong MgII-FeII
criterion has since been modified using our new HST Cycle 11 survey -
see below. Meanwhile, using the same technique, Lane (2000) surveyed
radio loud quasars for 21 cm absorption corresponding to 62 MgII
systems with the  WSRT, and derived a
similar result for $\Omega_{DLA}$ at $z<1.65$ (see  Lane \& Briggs 2001).

With the number of quasar UV spectra increasing rapidly in the HST
archives, it became possible to conduct a UV survey for MgII at
redshifts $0<z<0.15$.  Churchill (2001) used 147 HST archival
quasar/AGN spectra  to obtain a total redshift search path of 18.8,
and found 4 strong MgII systems. For $W_0^{\lambda2796} > 0.6$ \AA, he
derived $dn/dz=0.22^{+0.12}_{-0.09}$ at $\left<z\right>=0.06$, consistent with
the extrapolation of $dn/dz$ to lower redshifts for $W_0^{\lambda2796}
> 0.6$ \AA\ from Nestor (2004). All four quasars now have HST
UV spectra that include redshifted Ly$\alpha$ at the absorber
redshifts. Three of the  four have sub-DLA column densities  and no
information is available  for the fourth due to an intervening Lyman
limit system (Churchill 2001;  Rao HST GO program 9382; Keeney et
al. 2005). Thus, the number of  DLAs in this sample is $\le 1$. This
gives a 1$\sigma$  upper limit to the DLA redshift number density of
$n(z=0.06) \le 0.19$. While this result does not offer  significant
constraints, it is, nevertheless, consistent with the data points at
$z=0$  and at low redshift (see \S3).

\section{New results at low redshift}

Further progress could only be made with a dramatic increase in 
sample size. The SDSS sample of quasars, which numbered in the
thousands when this phase of our MgII-DLA project began, presented an
unprecedented leap in the  number of available survey quasars. The
previous largest MgII survey by  Steidel and Sargent (SS92) used
a sample of 103 quasars; the SDSS  Early Data Release included nearly
4000. Nestor (2004) used  SDSS-EDR quasar spectra  to search
for MgII systems with the aim of quantifying the statistical
properties of a large MgII sample (Nestor, Turnshek, \& Rao 2005,
henceforth NTR05) and to conduct follow-up work to search for DLAs. In
Cycle 11 (PID 9382), we targeted a sample of 83 MgII systems with
$W_0^{\lambda2796}\gtrsim 1$ \AA\ in 75 SDSS quasars. There were an
additional 16 weaker MgII systems observable in the same set of
spectra. Overall, useful UV information was obtained for 88 systems,
25 of which are DLAs.  For details of the sample and the selection
criteria used to include MgII systems in the survey, the reader is
referred to RTN05.  In total, we now
have 197 $z<1.65$ MgII systems with $W_0^{\lambda2796} > 0.3$ \AA\ for
which UV observations of the Ly$\alpha$ line have been obtained.
Forty one of these are DLAs.

\subsection{Metal line and HI correlations}

\begin{figure}
\centerline{\scalebox{0.5}{ \includegraphics{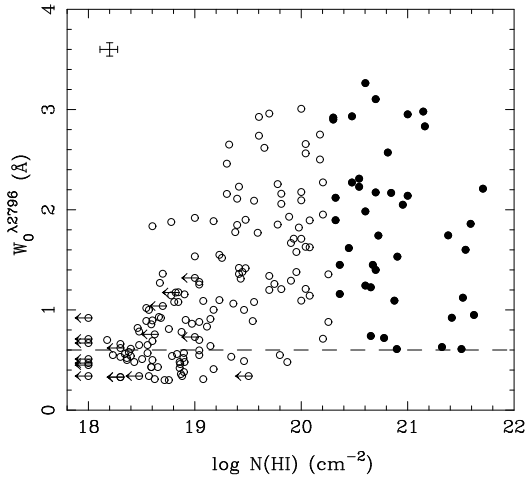}}}
\caption{Plot of \Wmi\ vs. $\log N(HI)$ from RTN05. Filled circles are
DLAs with $N(HI)\ge 2\times 10^{20}$  cm$^{-2}$. Arrows are upper
limits in $N(HI)$.  Typical uncertainties are given by the error bars
in the top left corner.}
\end{figure}

\begin{figure}
\centerline{\scalebox{0.5}{ \includegraphics{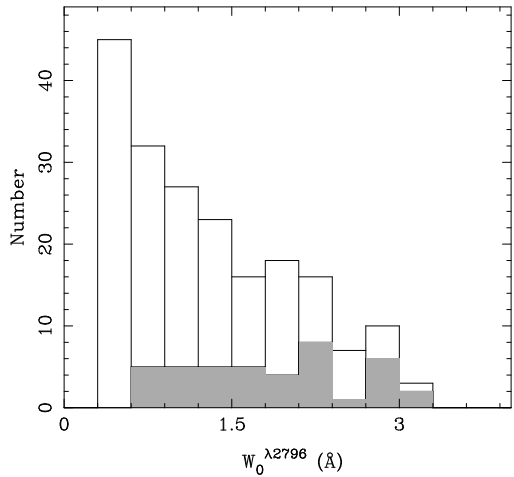}}}
\caption{Distribution of MgII $\lambda 2796$ rest equivalent widths,
\Wmi. The gray histogram represents systems that are DLAs. Note that
there are no DLAs in the first bin, i.e., for MgII \Wmi$<0.6$ \AA. The
number of DLAs is roughly constant for \Wmi$>0.6$ \AA, implying that
the fraction of MgII systems that are DLAs  increases with increasing
\Wmi. See RTN05.}
\end{figure}

Since the systems in our sample were selected based on the rest
equivalent  width of MgII $\lambda2796$, $W_0^{\lambda2796}$ and
$N(HI)$ measurements exist for all 197 systems. MgII $\lambda2803$,
the weaker member of the doublet, was also measured for all systems,
while the FeII $\lambda2600$ and MgI $\lambda 2852$ lines were
measured for only a subset.  Here, we explore correlations among metal
line rest equivalent widths and HI column density.  Figure 2 is a plot
of \Wmi\ versus $\log N(HI)$.  We note that the upper left region of
the figure is not populated implying that systems with \Wmi$>2.0$ \AA\
always have HI column densities $N(HI) > 1\times 10^{19}$ atoms
cm$^{-2}$. Figure 3 gives the distribution of MgII \Wmi; the DLAs form
the gray histogram. It is noteworthy that there are no DLAs with
\Wmi$<0.6$ \AA.\footnote{Only one known DLA has lower metal-line rest
equivalent  widths. The 21 cm absorber at $z=0.692$ towards 3C 286 has
\Wmi$=0.39$ \AA\ and  \Wf$=0.22$ \AA\ (Cohen et al. 1994). None of the
21 cm absorbers are included in  our analysis here because they are
biased systems (see RT00).}  In addition, the number of DLAs remains
roughly constant for \Wmi$>0.6$ \AA.  As a consequence, the fraction
of systems that are DLAs increases with  increasing \Wmi. This is
shown as a histogram in Figure 4; the y-axis  on the left gives the
fraction of DLAs as a function of \Wmi.  We also plot the mean HI
column density in each bin as solid circles with the scale shown on
the right. Upper limits are assumed to be detections. The left panel
includes all observed MgII systems and the right panel includes only
the DLAs. The vertical
error bars are standard deviations in the  mean and are due to the
spread of $N(HI)$ values in each bin, and the horizontal error bars
indicate bin size. For the MgII systems, there is a dramatic increase 
of a factor of $\approx 36$ in
the mean HI column density from the first to the second bin, beyond
which $\left<N(HI)\right>$ remains constant within the errors. For
systems with 0.3 \AA\ $\le$ \Wmi $<$ 0.6 \AA,  $\left<N(HI)\right> =
(9.7\pm 2.2)\times 10^{18}$ cm$^{-2}$, and $\left<N(HI)\right> = (3.5
\pm 0.7) \times 10^{20}$ cm$^{-2}$  for systems with \Wmi $\ge$  0.6
\AA. The right panel shows a trend for decreasing DLA column density
with \Wmi. The reasons for this are not obvious, but are likely to 
be due to small number statistics (see Figure 2), a real physical effect
or a selection effect that is not yet understood (Turnshek et al. 2005).

\begin{figure}
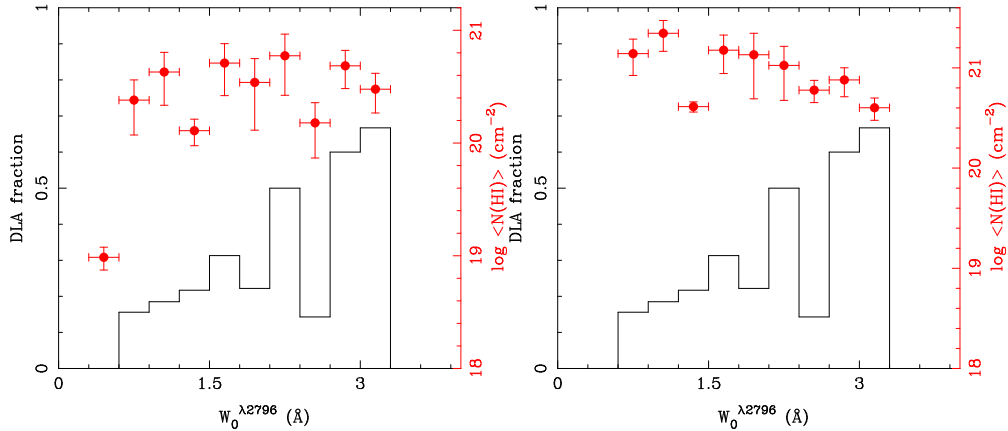

\centerline{\scalebox{0.45}{\includegraphics{rao04a.ps}\hfil
\includegraphics{rao04b.ps}}}
\caption{Both histograms show the fraction of MgII systems that are
DLAs as a function of MgII \Wmi, with the scale shown on the left
axis. The  solid circles are the logarithm of the mean HI column
density in each bin for all observed MgII systems in the left panel
and only for the DLAs in the right panel. The scale is shown on the right.}
\end{figure}

Figure 5a is a plot of  \Wmi\ vs. \Wf\ for systems with measured
values of \Wf, including upper limits.  In RT00 we found that 50\% of
the 20 systems  (excluding  upper
\begin{figure}
\centerline{\scalebox{0.5}{ \includegraphics{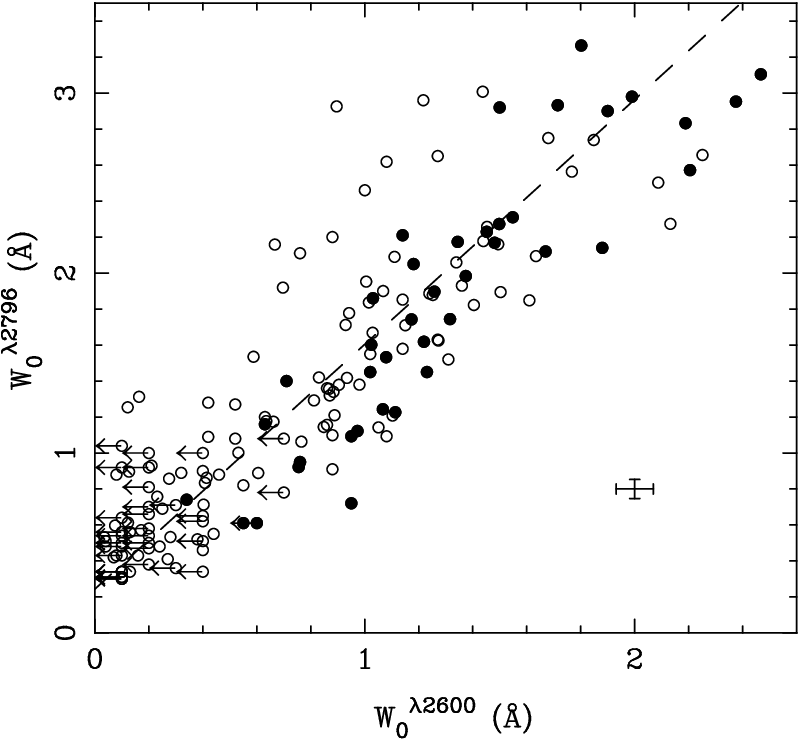}\hfil
\includegraphics{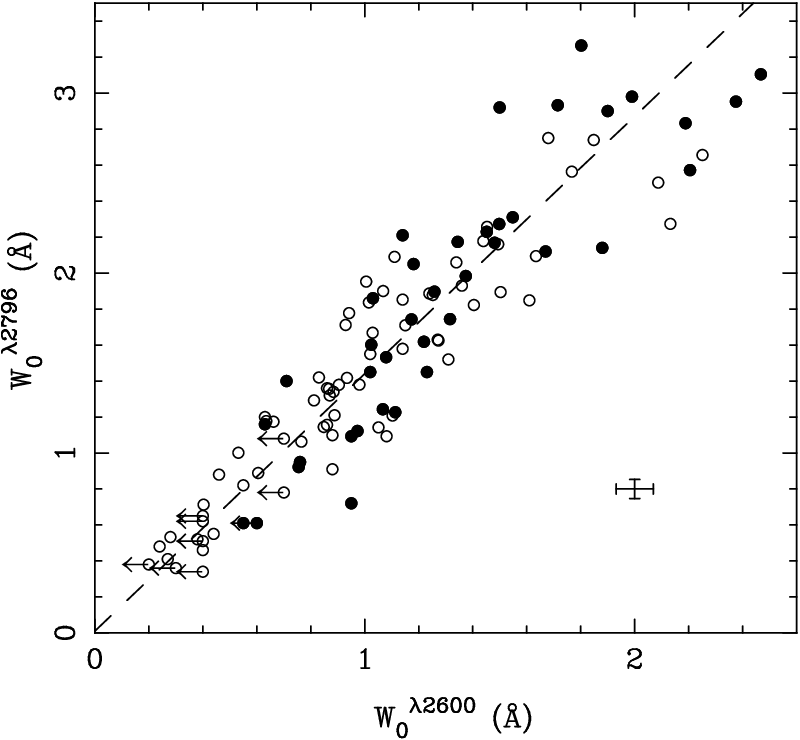}}}
\caption{{\it Left, a:} Plot of \Wmi\ vs. \Wf\ for all MgII systems
that have measured values of \Wf. Filled circles are DLAs. Typical
error bars are shown in the lower right corner.  The dashed line is
the best fit linear correlation with slope
$b=1.36$.  {\it Right, b:} Plot of \Wmi\ vs. \Wf\ for systems with
\Wmi/\Wf$<2$.  Non-DLAs in the upper left and
lower left  regions of the diagram have been eliminated and the
correlation is tighter than what is seen in Figure 5a. Also, 38\% of
all systems are DLAs regardless of  the values of \Wmi\ and \Wf; we
believe this to be a more robust predictor of the presence of DLAs in
MgII-FeII systems. The slope of the best fit linear correlation is
$b=1.43$. Results are from RTN05.}
\end{figure}
limits and 21 cm absorbers) with both \Wmi\ $> 0.5$ \AA\ and \Wf\ $>
0.5$ \AA\ are DLAs. Now, with the expanded sample that includes 106
systems in this   regime, we find that 36\% are DLAs. The dashed line
is a least-squares fit  with slope $b=1.36$. We note that DLAs do not
populate the  top left region of the diagram where the \Wmi\ to \Wf\
ratio is  $\gtrsim 2$. In fact, if the sample is restricted to systems
with \Wmi/\Wf$<2$,  all but one of the DLAs from Figure 5a are
retained, the outliers in the top left region are excluded as are most
systems in the lower left corner of the plot. Figure 5b shows this
truncated sample; the slope of the least-squares fit does not change
significantly. We find $b=1.43$. The only DLA that has been eliminated
is the one with the smallest value of  \Wf. However, given the
measurement errors for this system, its  \Wmi/\Wf ratio is within
1$\sigma$ of 2. This also implies that a system with metal line ratio
\Wmi/\Wf $>2$ has nearly zero probability of being a DLA.  For this
truncated sample with \Wmi/\Wf$<2$, but no restrictions on the
individual values of \Wmi\ or \Wf,  38\% are DLAs.  In addition, all
known 21 cm absorbers, including the $z=0.692$ system towards 3C 286
mentioned above, have \Wmi/\Wf$<2$. Thus, the \Wmi/\Wf\ ratio is a
more robust predictor of the presence of DLAs.

This result is shown more dramatically in Figure 6. The ratio
\Wmi/\Wf\ is plotted as a function of $N(HI)$ in Figure 6a.  Ratios
above 5 are not shown for clarity. These are mainly confined to $\log
N(HI) < 19.2$ with  only one system above this column density at $\log
N(HI) = 19.6$.  The DLAs populate the region of the plot where
$1\lesssim$\Wmi/\Wf$\lesssim 2$; the two outliers lie within $1\sigma$
of this range.  A plot of the ratio \Wmi/\Wf\ vs. MgI \Wmiii\ for
systems with  measured values of  \Wmiii, including upper limits, is
shown in Figure 6b. Again, the DLAs are  confined to the region where
$1\lesssim$\Wmi/\Wf$\lesssim 2$, but span the entire range of
\Wmiii. The two systems outside the range $1\lesssim$\Wmi/\Wf$\lesssim
2$ from Figure 6a do not have any information on \Wmiii. We find that
9 out of the 11 systems with \Wmiii$>0.8$ \AA\  are DLAs.  We also note
that systems with \Wmi/\Wf$\gtrsim2$ are likely to have  low values of
\Wmiii.

\begin{figure}
\centerline{\scalebox{0.5}{ \includegraphics{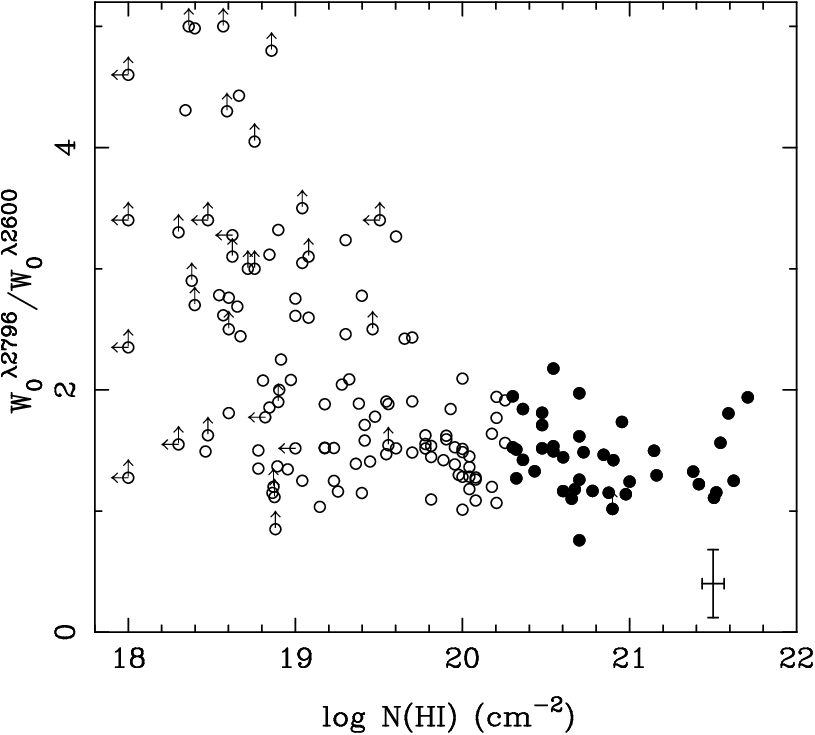}\hfil
\includegraphics{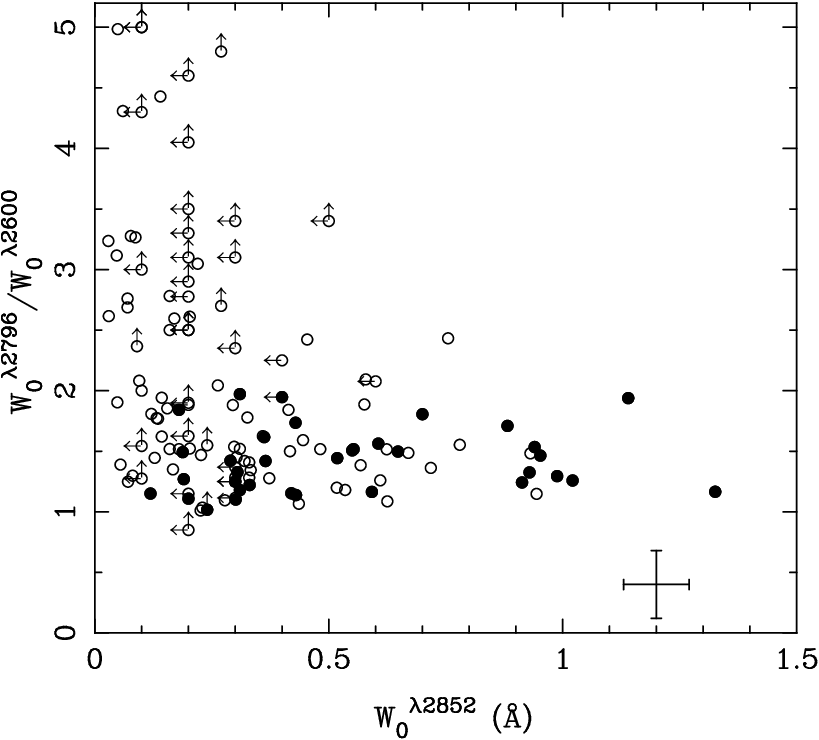}}}
\caption{{\it Left, a:} Plot of \Wmi/\Wf\ vs. $N(HI)$. Systems with
\Wmi/\Wf$>5$ are not shown for clarity; all of these have $\log
N(HI)<19.6$. The DLAs are confined to the region of the plot where
$1\lesssim$\Wmi/\Wf$\lesssim 2$.  {\it Right, b:} Plot of \Wmi/\Wf\
vs. MgI \Wmiii. Solid circles are DLAs. Typical error bars are shown
in the lower right corner. Again, the DLAs are  confined to the region
where $1\lesssim$\Wmi/\Wf$\lesssim 2$, but span the entire range of
\Wmiii. }
\end{figure}

For completeness, we also plot \Wf\ vs. $\log N(HI)$ in Figure 7a and
\Wmiii\ vs. $\log N(HI)$ in Figure 7b. There is no obvious trend in
these distributions except for the fact that the upper left regions of
the plots are not populated. There are no high rest equivalent width,
low HI column density systems. This is not a selection effect since
column densities as low as $10^{18}$ cm$^{-2}$ can often be easily
measured.  As is the case for \Wmi, this implies that systems  with
\Wf $ \gtrsim 1$\AA\ or \Wmiii $\gtrsim 0.5$\AA\  generally have HI
column densities $N(HI) >10^{19}$ cm$^{-2}$.  Below this fairly
sharp boundary, metal-line rest equivalent widths  span all values of
HI column density.

\begin{figure}
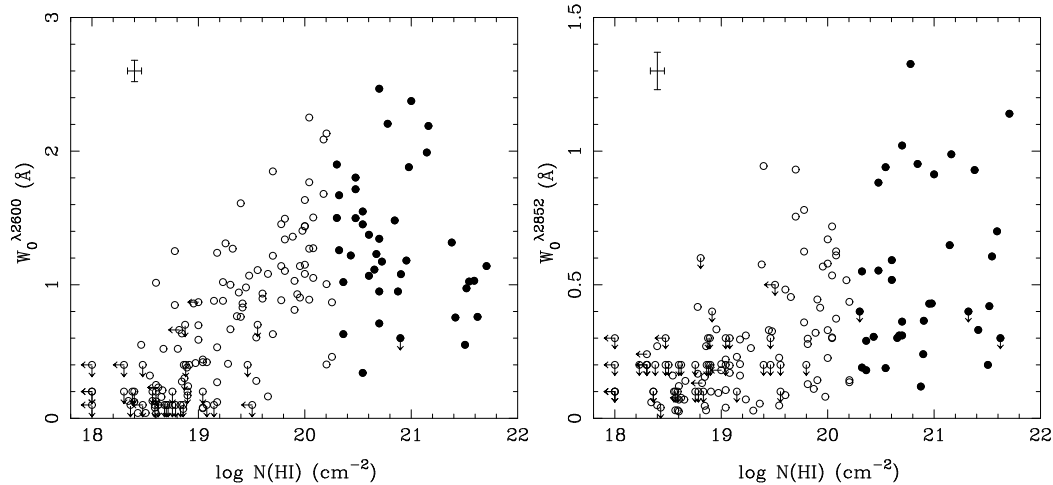

\centerline{\scalebox{0.5}{ \includegraphics{rao07a.ps}
\includegraphics{rao07b.ps}}}
\caption{{\it Left, a:} Plot of \Wf\ vs. $\log N(HI)$. Arrows indicate
upper limits.  Typical uncertainties are given by the error bars in
the top left corner.  {\it Right, b:} Plot of \Wmiii\ vs. $\log
N(HI)$. Arrows indicate upper limits.  Typical uncertainties are given
by the error bars in the top left corner.}
\end{figure}

\subsection{Interpretation of Absorption Characteristics}

How can these trends be interpreted? Apart from the upper envelopes in
Figures 2 and 7, there is no correlation between metal-line rest
equivalent width and HI column density. Since the metal lines are
saturated, the rest equivalent width is more a measure of kinematic
velocity spread and not column density. High resolution observations
of MgII absorption lines have shown that the stronger systems break up
into many components (e.g., Churchill et al. 2003), and span velocity
intervals of up to 400 km/s. Turnshek et al. (2005) show line
equivalent widths in velocity units of $\gtrsim800$ km/s  in the
strongest systems found in the SDSS.  These highest equivalent width
systems  may arise in galaxy groups; however, the more common systems
like those in our DLA survey are more likely to arise in clouds that
are  bound in galaxy-sized potentials. A   DLA is observed if one of
the clouds along the sightline happens to be  cold (less than a few
100 K), dense, and with a velocity dispersion of a few 10s of km/s.
The more the number of clouds along the sightline, the higher the
probability of encountering a DLA. This would explain the higher
fraction of DLAs among large \Wmi\ systems and the lack of a
correlation between  \Wmi\ and $N(HI)$. Only very rarely would a
sightline intersect a single cloud resulting in small \Wmi\ {\it and}
high $N(HI)$, as in the 3C 286 system  described in \S3.1. This
probabilistic approach to explain metal-line and  HI strengths in
high-$N(HI)$ absorbers was also proposed by Briggs \& Wolfe (1983) to
explain their MgII survey for 21 cm absorbers. They proposed a
two-phase model  where the 21 cm absorption is produced in galaxy
disks, and the metal-line  components that do not produce 21 cm
absorption are produced in galactic halos.  However, this
multi-component/cloud model could be valid in any gas-rich galaxy, as
is evidenced by DLA galaxy imaging studies (Le Brun et al. 1997;
Rao \& Turnshek 1998; Turnshek et al. 2001; Rao et al. 2003; Turnshek
et al. 2004). The disk models of Prochaska \& Wolfe (1997) and the
Haehnelt et  al. (1998) models of infalling and merging clouds could
reproduce these observations equally well. In other words, DLAs arise
in pockets of cold gas   embedded within warm diffuse gas or gas
clouds in any bound system.
 
Twenty one cm observations of low-redshift DLAs also reveal some cloud
structure. For example, the $z=0.313$ system towards PKS 1127$-$145
shows 5 components and the $z=0.394$ system towards  B0248+430 is
resolved into 3 components (Lane 2000; Lane \& Briggs 2001; Kanekar \&
Chengalur 2001). Since the MgII line for these systems has not been
observed at a resolution as high as the 21 cm observations, a
one-to-one correspondence between the metal-line and 21 cm  clouds
cannot be drawn. In other instances, both warm and cold gas have been
detected in a 21 cm DLA; Lane et al.  (2000) find that two-thirds of
the column density in the $z=0.0912$ DLA towards B0738+313 is
contained in warm phase gas, and the rest is contained in two narrow
components. The $z=0.2212$ absorber towards the  same quasar was also
found to exhibit similar characteristics (Kanekar  et al. 2001). In
each of these cases, the line of sight probably intersects two cold
clouds in addition to warm diffuse gas spread over a wider range of
velocities that can be detected only in 21 cm  observations of very
high sensitivity. There are also several instances of DLAs not being
detected at 21 cm (Kanekar \& Chengalur 2003). High spin temperatures
($T_s>1000$ K) corresponding to warm diffuse gas and/or covering
factors less than unity towards extended quasar radio components have
been  suggested as possible explanations (Kanekar \& Chengalur 2003;
Curran et al. 2005).

Clearly, a wide variety of cloud properties and their combinations are
responsible for the observed properties of MgII, DLA, and 21 cm
absorption lines.  Observations in the 21 cm line have the added
complication that the background radio and optical sources may sample
different sightlines, thus making interpretation difficult. DLA
observations, on the other hand, do not suffer from this  effect.
Large simulations of galaxy sightlines with varying cloud properties
that reproduce the metal-line, DLA correlations shown in Figures 2-7
should be the next step towards improving our understanding of these
absorption line systems. The simulations should not only reproduce the
frequency of occurrence of DLAs in MgII systems, but also the  number
density evolution of MgII systems and DLAs.

\subsection{The redshift number density of DLAs, $n_{DLA}$}

The redshift number density of DLAs, $n_{DLA}$, sometimes written as
$dn/dz$,  can be determined using the equation
\begin{equation}
n_{DLA}(z) = f(z)\, n_{MgII}(z),
\end{equation}
where $f(z)$ is the fraction of DLAs in a MgII sample as a function of
redshift and $n_{MgII}(z)$ is the redshift number density of MgII
systems.  With the systems detected in SDSS EDR quasar spectra, Nestor
(2004) derived
\begin{equation}
dn/dz=N^*\,(1+z)^\alpha\,e^{-\frac{W_0}{W^*}(1+z)^{-\beta}},
\end{equation}
where we have retained his notation and $W\equiv$ \Wmi\ (see also
NTR05).  $N^*$, $W^*$, $\alpha$, and $\beta$  are constants.  This
expression is an integral over all \Wmi\ greater than $W_0$.  Since
our MgII sample was assembled under various selection criteria that
were modified based on our initial results (RT00), $n_{DLA}(z)$ had to
be evaluated carefully. Again, we refer the reader to RTN05 for
details on how $n_{DLA}(z)$ was determined.

Figure 8 shows the results for $n_{DLA}(z)$ at low redshift split into
two redshift  bins (solid squares). We find 18 DLAs in 104 MgII
systems in the redshift interval $0.11<z\le0.9$ with $n_{DLA}(z=0.609)
= 0.079 \pm 0.019$ and 23 DLAs in 94 MgII systems in the redshift
interval $0.9<z\le1.65$ with $n_{DLA}(z=1.219) = 0.120  \pm 0.025$.
The points are plotted at the mean redshift of the MgII samples.  The
high-redshift data points are from Prochaska \& Herbert-Fort (2004)
and the $z=0$ point was estimated by Zwaan (in preparation) from a
WSRT survey of HI in the local universe. The solid curve in Figure 8a
is a no-evolution curve in the standard $\Lambda$CDM cosmology, that
we refer to as the ``737'' cosmology, where
($h$,$\Omega_M$,$\Omega_\Lambda$) = (0.7, 0.3, 0.7).  This curve,
which  has been normalized at the $z=0$ data point, shows that the
comoving  cross-section for absorption declined rapidly by a factor
of $\approx 2$ until $z\approx 2$ and has remained constant since
then. This behavior might be a consequence of what has been observed
in other studies of galaxy evolution, namely, that today's galaxies
were in place by $z\approx 1$ and are a consequence of  rapid merger
and/or collapse events that occurred prior to this epoch.

\begin{figure}
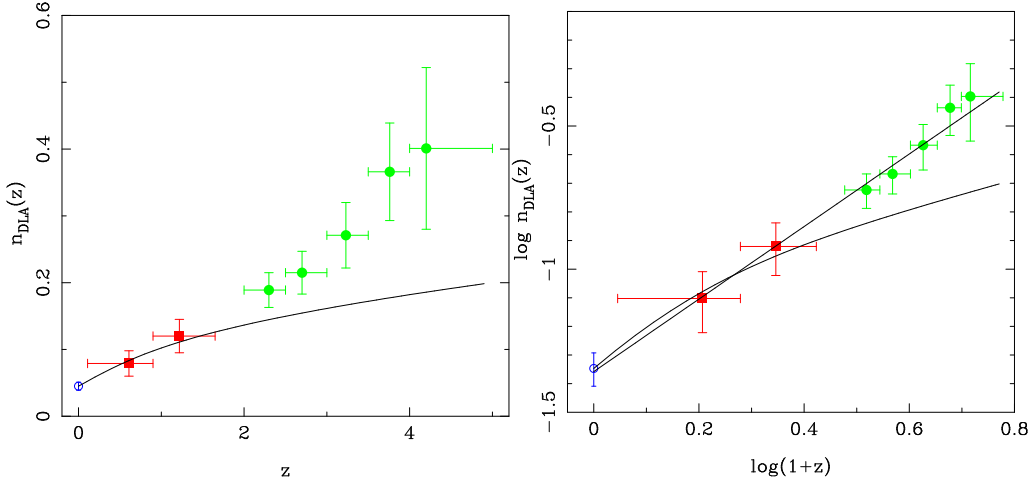

\scalebox{0.5}{ \includegraphics{rao08a.ps}\hfil
\includegraphics{rao08b.ps}}
\caption{{\it Left, a:} Plot of $n_{DLA}(z)$ versus redshift. The new
low-redshift data points from RTN05 are shown as filled squares. The
high-redshift points (filled circles) are from Prochaska \&
Herbert-Fort (2004) and the $z=0$ data point is from an analysis of
local HI  (Zwaan, in preparation). The solid line is a no-evolution
curve in the ``737'' cosmology with
($h$,$\Omega_M$,$\Omega_\Lambda$) = (0.7, 0.3, 0.7)  normalized at 
$z=0$. The comoving cross-section for absorption declined rapidly by a
factor of $\approx 2$ until $z\approx 2$ and has remained constant
since then. {\it Right, b:} Plot of $\log n_{DLA}(z)$ as  a
function of $\log(1+z)$. The straight line is the power law fit to
the data points with slope $\gamma=1.27\pm 0.11$, and the  curve is
the no-evolution function shown in Figure 8a.}
\end{figure}

It has been customary in quasar absorption line studies to plot the
logarithm of the redshift number density in order to illustrate its
power law dependence with redshift, i.e.,  $n(z)=n_0(1+z)^\gamma$. In
$\Lambda=0$ cosmologies, the exponent is a  measure of the evolution
in this quantity. For example,  $\gamma=1$ for $q_0=0$ or $\gamma
=0.5$ for $q_0=0.5$ imply no intrinsic evolution of the absorbers, and
any significant departure from these values for $\gamma$ is evidence
for evolution in the comoving number density of absorbers. We plot
$\log n_{DLA}(z)$  as a function of $\log (1+z)$ in Figure 8b. The
straight line is the power law fit to the data points with slope
$\gamma=1.27\pm 0.11$, and the  curve is the same no-evolution
function  shown in the left panel. Thus, in the past the observations
would have been interpreted as being consistent with the DLA absorbers
undergoing no intrinsic evolution in a $q_0=0$ universe, and
marginally consistent  with evolution in a $q_0=0.5$ universe. With
the now widely accepted concordance cosmology, the interpretation has
changed quite dramatically; as noted above, the nature of the
evolution is redshift dependent.
 
The possibility of a dust bias in DLA statistics at low redshift was
addressed by Ellison et al. (2004) using the same CORALS survey that
Ellison et al. (2001) used to show that the effect of dust at
$z\approx 2$ was minor. They determined the MgII $dn/dz$ for
redshifts $0.6<z<1.7$, and found excellent agreement with the results
of NTR05 who used the SDSS-EDR quasar sample. Based on the RT00 strong
MgII-FeII selection for DLAs, they concluded that dust  bias was not
important for DLAs at low redshift either.

\subsection{The cosmological mass density $\Omega_{DLA}$}

We can determine $\Omega_{DLA}$ from the DLA column densities  and
$n_{DLA}(z)$ via the expression
\begin{equation}
\Omega_{DLA}(z)= \frac{\mu m_H H_0}{c \rho_c} n_{DLA}(z)
\left<N(HI)\right>  \frac{E(z)}{(1+z)^2},
\end{equation}
where
\begin{equation}
E(z)=(\Omega_M(1+z)^3 + (1-\Omega_M-\Omega_\Lambda)(1+z)^2 +
\Omega_\Lambda)^{1/2}.
\end{equation}
Again, the ``737'' cosmology has been used in the calculation of
$\Omega_{DLA}$.  Also, $\mu=1.3$ corrects for a neutral gas
composition of 75\% H and 25\% He by mass, $m_H$ is the mass of the
hydrogen atom,  $\rho_c$ is the critical mass density  of the
universe, and $\left<N(HI)\right>$ is the mean HI column density of
DLAs in each redshift bin.

\begin{figure}
\centerline{\scalebox{0.5}{ \includegraphics{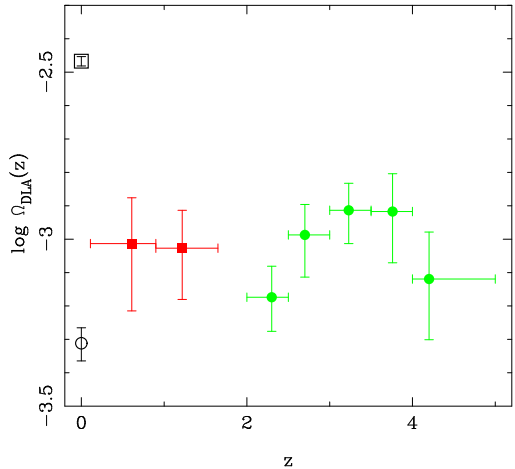}}}
\caption{Cosmological mass density of neutral gas, $\Omega_{DLA}$, as
a  function of redshift. The filled squares are the new low-redshift
data points from RTN05. The high-redshift points (filled circles) are
from Prochaska   \& Herbert-Fort (2004) and the open circle at $z=0$
is from Zwaan et al.  (2005). The open square at $z=0$ is the mass
density in stars estimated by Panter et al. (2004) from SDSS data.
While the statistics have improved considerably,
our basic conclusion from RT00 has remained unchanged, namely, that
the cosmological mass density of neutral gas has remained constant
from $z\approx 5$ to $z \approx 0.5$.}
\end{figure}
In contrast to the redshift number density evolution shown in Figure
8, we find  that $\Omega_{DLA}$ has remained constant from $z\!=\!5$
to $z\!=\!0.5$ to within the uncertainties. Figure 9 shows the new
results as solid squares.  Specifically, for the redshift range
$0.11<z\le 0.90$, we find $\Omega_{DLA}(z\!=\!0.609)\!=\!(9.7 \pm 3.6)
\times 10^{-4}$ and for the range $0.90\!<\!z<\!1.65$ we get
$\Omega_{DLA}(z\!=\!1.219)=(9.4 \pm 2.8) \times 10^{-4}$.  The
uncertainties have been reduced considerably in comparison to our
results in RT00.  The reasons for this are two fold. First, the
uncertainty in $n_{MgII}$ has been significantly reduced due to the
fact that the MgII sample size was increased 10-fold. Second, the
number of DLAs in each bin has increased by more than a factor of
3. Thus, the uncertainties in the low- and high-redshift data points
are now comparable. Note  that the statistics of the high-redshift
data are also improved due to the inclusion  of an SDSS DLA sample
(Prochaska \& Herbert-Fort 2004). Nevertheless, our basic conclusion
from RT00 has remained unchanged, namely, that the cosmological mass
density of neutral gas remains roughly constant from $z\approx 5$ to
$z \approx 0.5$.

The drop in redshift number density from $z=5$ to $z=2$ along with a
constant mass density in this range indicates that while the product
of galaxy cross-section and co-moving number density  is declining,
the mean column density per absorber is increasing. This is, again,
consistent with the assembly of higher density clouds as galaxy
formation  proceeds.

On the other hand, a constant cross-section from $z\approx 1$ to $z=0$
along with  a drop in mass density from $z\approx 0.5$ to $z=0$ is
indicative of star formation  that depletes the highest column density
gas while keeping the absorption  cross-section constant. This would
in turn require that the column density distribution of DLAs change
such that the ratio of high to low column densities decreases from
low-redshift to $z=0$. As we will see in the next section, the column
density distribution does show some evidence for this.

\subsection{The column density distribution $f(N)$}

\begin{figure}[h]
\centerline{\scalebox{0.5}{ \includegraphics{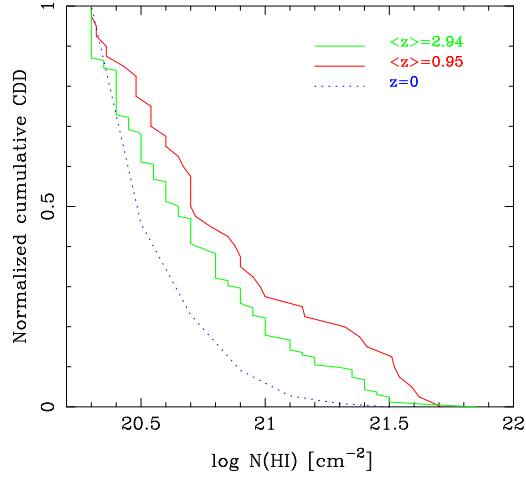}}}
\caption{The normalized cumulative CDD of DLAs
for three redshift regimes. The red, thick solid line
includes the 41 low-redshift DLAs at $\left<z\right>=0.95$ (RTN05).
The green, thin solid line includes 163 high-redshift
DLAs at  $\left<z\right>=2.94$ (Prochaska \& Herbert-Fort 2004),  and
the blue, dotted curve is the $z=0$ estimate from Ryan-Weber et
al. (2003, 2005).}
\end{figure}
\begin{figure}
\centerline{\scalebox{0.5}{ \includegraphics{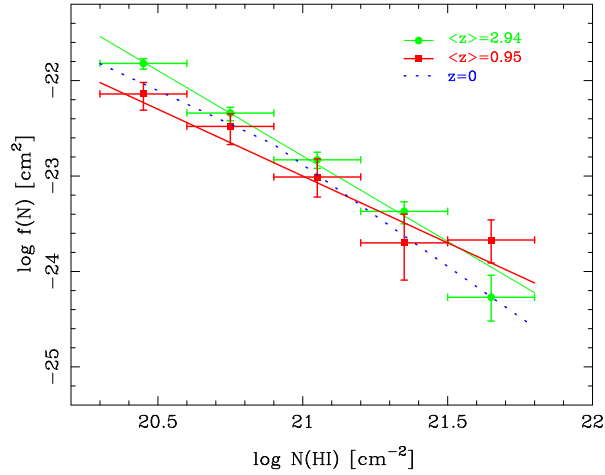}}}
\caption{Absolute CDD function for the
three redshift regimes. The red, thick solid line is a least squares
fit to the low-redshift data points (RTN05) and has slope $\beta=1.4$;
the green, thin solid line is a least squares fit to the high-redshift
data points (Prochaska \& Herbert-Fort 2004) with
$\beta=1.8$.  The dotted line is the CDD with $\beta = 1.4$ for 
$\log N(HI)<20.9$ and $\beta=2.1$ for $\log N(HI)\ge 20.9$
(Ryan-Weber et al. 2003, 2005).}
\end{figure}

Figure 10 shows the normalized cumulative column density distribution
(CDD) for the three redshift regimes. The dotted curve is the $z=0$
distribution derived by Ryan-Weber et al. (2003, 2005) from  HIPASS
data. The red (thick, solid) curve is derived from the 41 DLAs at low
redshift (RTN05) and the green (thin, solid) curve  is from the
``total'' sample of Prochaska \& Herbert-Fort (2004). The  change in
the three CDDs with redshift  is exactly what is expected based on the
$n_{DLA}$ and $\Omega_{DLA}$ results.  Namely, that the low-redshift
CDD shows a higher incidence of high column density systems than at
high redshift presumably due to the assembly of gas as galaxy
formation proceeds, followed by a decrease in  the fraction of high
column density systems to $z=0$, presumably due to the  depletion of
gas during star formation. Thus, at least qualitatively, the
evolutionary behavior of $n_{DLA}$, $\Omega_{DLA}$, and the CDD are
entirely consistent with one another. A KS test shows that there is a
25\% probability that the high- and low-redshift curves are drawn from
the same population; this is significantly higher than what we
observed in RT00, where the  two samples had only a 2.8\% probability
of being drawn from the same population. However, the general trend
that the low-redshift sample has a  higher fraction of high column
density system still remains.

Figure 11 is a plot of the log of the absolute CDD function, $\log
f(N)$, as a  function of $\log N(HI)$ for the three redshift regimes.
The turnover with redshift is most apparent in  the lowest and highest
column density bins.  We derive $\beta = 1.4\pm 0.2$ and $\beta=1.8 \pm 0.1$
at low and high redshift respectively, where the CDD is expressed as 
$f(N)=BN^{-\beta}$. At
$z=0$, Ryan-Weber et al. (2003) derive $\beta=1.4\pm0.2$ for $\log
N(HI)<20.9$ and $\beta=2.1\pm0.9$ for $\log N(HI) \ge 20.9$. The general
form of the absolute CDD does not vary considerably  with redshift,
which in turn explains the roughly constant value of $\Omega_{DLA}$.
The differences in the $f(N)$ distributions are subtle, implying that
the gas content in DLAs is not changing drastically. This is strong
evidence that DLAs do not have high SFRs and are,
therefore, a different population of objects than those responsible
for much of the observed luminosity in the high redshift universe.  On
the other hand, a non-evolving DLA population might be observed if the
gas that is used up in star formation is replenished from the
inter-galactic medium at a comparable rate. This possibility seems
rather contrived, and requires more proof than the current
observational evidence can provide. In the next section, we discuss
further evidence that  suggests that DLAs and Lyman break galaxies
(LBGs) are mutually exclusive  galaxy populations.

\section{The Nature of DLAs}

With the identification of DLAs in quasar spectra, follow-up work
involving the study of DLA galaxies, their environments, DLA gas
metallicities, kinematics, temperatures, and ionization conditions, and
numerical and semi-analytic modelling has kept many astronomers busy
for decades. A keyword search for ``damped Ly'' in NASA's abstract
database returns nearly 1000 results, too many to reference
here. Despite this, a consensus on the nature of these objects is
still lacking. One of the primary reasons for this is that the
selection effects inherent in the observations are still being
sorted out.  Many of these issues were raised at this conference, and
we address some of them here. See also Turnshek et al. (2005, these 
proceedings) for further discussion on DLA selection effects.

The conclusion that DLA surveys identify the bulk of the neutral gas
in the universe is based on three results or assumptions. First,
integration of  the CDD  shows that a relatively small fraction of the
neutral gas is contributed by Lyman limit and sub-DLA absorption
systems with $3\times10^{17} < N(HI) < 2\times10^{20}$  cm$^{-2}$, at
least  for $z<3.5$ (P\'eroux et al. 2003), and perhaps at all
redshifts (Prochaska \& Herbert-Fort 2004). Second, as discussed
earlier,  dusty regions do not cause DLA surveys to miss a large
fraction of the neutral gas (Ellison et al. 2001, 2004).  Third, the
biases introduced by gas cross section selection are small. The
important points to emphasize here are that the interception (or
discovery) probability is the product of gas cross section times
comoving absorber number density, and no DLAs with $N(HI) >
8\times10^{21}$ cm$^{-2}$ have been discovered. Thus, the third
assumption requires that rare systems with relatively low gas cross
section and very high HI column density are either absent or have not
been missed to the extent that the neutral gas mass density will be
significantly underestimated by quasar   absorption line
surveys. This third assumption might have to be  reevaluated
in order to explain the discrepancy between the  star formation
history (SFH) of DLAs as inferred from their HI column densities and
that determined from galaxies that  trace the optical luminosity
function. We  explored this connection in Hopkins, Rao, \& Turnshek
(2005) and summarize it here.

\begin{figure}
\centerline{ {\scalebox{0.3}{\rotatebox{270}
{\includegraphics{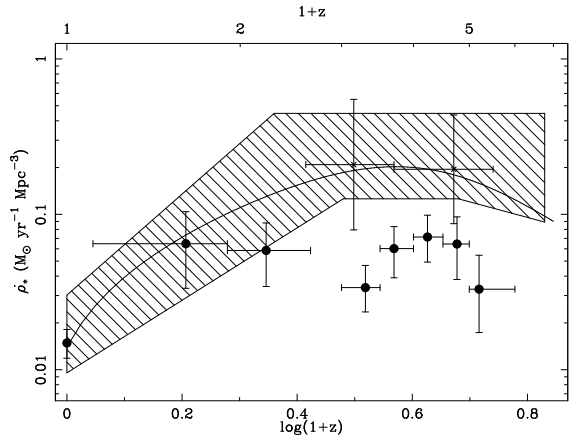}}}}}
\caption{ SFR density of DLAs (filled circles) as a function of
redshift from Hopkins, Rao, \& Turnshek (2005). Crosses are
estimates from Wolfe et al.  (2003) using the CII* method (but see 
footnote). The hatched
region encompasses the majority of SFR estimates in the literature
(Hopkins 2004) and the solid curve is a fit to these data.}
\end{figure}
We can use the global Schmidt law relating gas surface density and
SFR to infer the star forming properties of the
DLAs. Assuming that the Schmidt law, quantified by Kennicutt (1998),
is valid at the DLA redshifts we can convert DLA column densities into
gas surface densities and estimate their SFR. In Hopkins et al. (2005)
we show that the SFR density of DLAs relates to $n_{DLA}(z)$ and
$\Omega_{DLA}(z)$ as follows:
\begin{equation}
\dot{\rho}_* = 4.0 \times 10^{-15}
\left(\frac{dX/dz}{n_{DLA}}\right)^{0.4}
\left(\frac{\Omega_{DLA}}{\mu} \rho_{c}\right)^{1.4}
\end{equation}
where $\dot{\rho}_*$ is the SFR density in units of $\rm M_{\odot}$
yr$^{-1}$ Mpc$^{-3}$ and  $dX/dz$ is the absorption distance which
equals $(c/H_0)(1+z)^2/E(z)$.  $E(z)$, $\mu$, and $\rho_c$ are as
defined in \S3.4.  Figure 12 shows the SFR density of DLAs estimated
using equation 4.1 and  the values of $n_{DLA}(z)$ and
$\Omega_{DLA}(z)$ plotted in Figures 8 and 9.  The hatched region
encompasses the majority of SFR estimates in the literature (Hopkins
2004) and the solid curve is a fit to these data. The crosses are
estimates of  the SFR in DLAs using the CII* absorption line technique
of Wolfe et al. (2003).  The CII* results are consistent
with the fitted curve derived from the SFH of
luminous galaxies, and have been interpreted as  evidence that the DLA
population is the same as the LBGs\footnote{Using more recent 
UV background radiation estimates, Wolfe (2005, these proceedings) finds 
lower CII* estimates of DLA SFR density which are now 
in agreement with the Hopkins et al. result.}. 
However, the few high-redshift DLAs for
which host galaxies have been identified yield SFRs that are at the
very faint end of values determined for LBGs (e.g., Weatherley et
al. 2005,  M{\o}ller et al. 2002). Recent results from semi-analytic
models and numerical simulations  by Okoshi \& Nagashima (2005) and
Nagamine et al. (2004) indicate that the average masses of DLA
galaxies are likely to be low, around 10$^9$ M$_\odot$, which is
consistent with low  integrated levels of SFR. Okoshi \& Nagashima
(2005) find a broad range of SFRs from $10^{-6}$ to 100 M$_\odot$
yr$^{-1}$, with a mean value of 0.01 M$_\odot$ yr$^{-1}$ for DLAs at
redshifts $z<1$, and conclude that DLAs are typically dwarf systems
with low SFRs. Therefore, it is entirely possible that the high redshift
observations of Weatherley et al. (2005), and
M{\o}ller et al. (2002) are sampling the brightest end of DLA galaxy
properties. Lanzetta et al. (2002) have  also shown, through similar
HI surface density and SFR arguments that LBGs have SFR surface densities
that are up to 3 orders of magnitude higher than that of the DLAs. In
fact, their SFR densities do  not overlap for the
observed DLA redshift range $1.65<z<5$ (see Lanzetta  et al. figure
3b). This implies that the high SFR surface density objects have very
high column density neutral plus molecular gas ($10^{22}$ to $10^{25}$
cm$^{-2}$).  These are comparable to the surface gas and SFR
densities seen in local starburst nuclei (Kennicutt
1998). Observations of local star forming galaxies have shown that
$N(HI)$ is always  $\gtrsim 10^{-2} N(H_2)$ in the high SFR regions
(e.g., Wong \& Blitz 2002),  and therefore, could lie above the
currently observed DLA regime.  If these systems have very small
interception cross sections, they may be missed in surveys for
DLAs. For example, an absorber with a size of about 100 pc, comparable
to giant molecular clouds (GMCs) which are the sites of star
formation, has a cross section that is $\approx 10^4$  times smaller
than known DLAs, which typically have effective radii of  $\approx 10$
kpc (Monier et al. 2005). Assuming that there are on the order of 10 GMCs per galaxy, the
total cross section per unit volume, i.e., interception probability,
for these very high column density gas systems would be on the order
of $10^3$ times smaller.  This means that $10^3$ DLAs need to be
detected in order to find one very high column density system.  With
the SDSS, we are getting close, but are not quite there yet. A one in a
thousand system with $N(HI+H_2)=10^{24}$ cm$^{-2}$ would increase the 
SFR density of DLAs by more than a factor of 2, and bring the DLA SFR density
into agreement with the luminous SFR density. Searches
for molecular gas in DLAs have resulted in only a handful of
detections.  Moreover, the molecular gas fraction in the few DLAs with
H$_2$ detections  is very small (e.g. Ledoux et al. 2003), and  is
consistent with the  idea that the known sample of DLAs does not trace
the majority of the star forming gas in the universe.

A similar conclusion is reached when considering the low metallicities of
DLAs.  Figure 13a shows the most recent compilation of DLA
metallicities from  Rao et al. (2005b). Pettini and collaborators were
the first to make  metallicity measurements of a large number of DLAs;
the early data indicated that DLA metallicities do not evolve from redshifts
$z\approx 3.5$ to $z\approx 0.5$ (Pettini et al. 1999 and references
therein). With a much larger sample, Prochaska et al. (2003) and Rao
et al. (2005b), the latter with more measurements at $z<1.65$, showed
that DLA metallicities do show evidence for  evolution. In Figure 13b,
we show the metal mass density as a function of redshift. The DLA
metallicity data from Rao et al. (2005b) are converted to metal mass
by assuming $\log \Omega_{DLA} = -3$ and a solar metal mass fraction
of $Z_\odot=0.02$. The  metal mass density in stars, shown as the
hatched region in Figure 13b, can be derived from the SFR, since the
SFR is  related to the metal production rate (e.g., Madau et
al. 1996). Stellar population synthesis results indicate that
$\dot{\rho}_* = 63.7 \dot{\rho}_Z$ (Bruzual \& Charlot 2003, Conti et
al. 2003). The $z=0$ and $z=2.5$  results of Dunne et al. (2003) are
also shown. Dunne et al. used the dusty, high SFR,
submillimeter galaxy population at $z=2.5$ to derive  their result,
which is reasonably consistent with the evolution of $\rho_Z$ from the
SFH.  They concluded that high redshift DLAs are not the same
population as these star forming galaxies, which contain the majority
of the metals at this epoch.  The DLAs have about two orders of
magnitude less mass in metals than star forming galaxies. The
combination of low metallicity and low gas mass density naturally
leads to a space density of metals that is significantly lower than in
the luminous galaxy population. This ``missing metals'' problem cannot
be solved simply by invoking spheroids, dwarfs, low surface brightness
galaxies, or gas at large galactocentric distance. Instead, this problem 
may in fact be the result of missing a substantial  fraction of the
metal-enriched gas in DLA surveys, for example in the GMCs mentioned
above, due to their very small  interception probability. That
molecules are detected in DLAs with higher than average metallicities
is consistent with this idea (Ledoux et al. 2003).  Thus, the SFH
 and missing metals problem can both be attributed to
this same selection effect. All of this holds true at high redshift.
We note, however, that the SFR in DLAs and  luminous objects at $z<1$
are comparable, but their metal mass densities are not. This can be
explained by noting that by $z\approx1$, the SFR in  luminous, high
SFR, galaxies begins to decline, i.e., the luminosity density of the
universe fades, leaving behind the low SFR DLA  galaxies as the main
contributor to the volume averaged SFR. The metal mass density, on the
other hand, has been building up in the high SFR galaxies at high
redshift, and continues to remain high.  In fact, $\rho_Z$  plateaus
at $z\lesssim 1$ because the DLAs have low SFRs and do not  increase
the metal mass density significantly.

\begin{figure}
{\scalebox{0.27}{\rotatebox{270} {\includegraphics{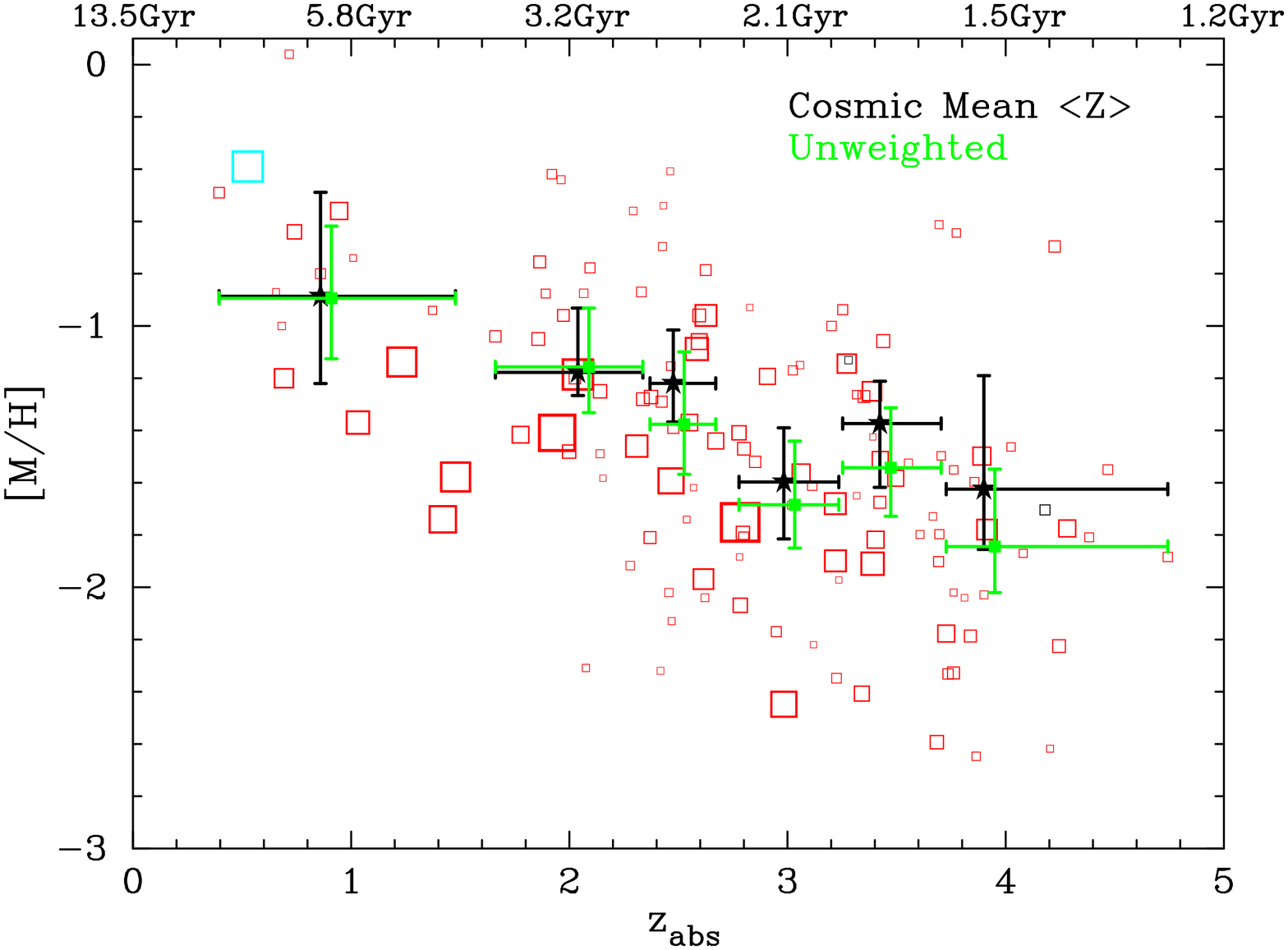}}}}
{\scalebox{0.27}{\rotatebox{270} {\includegraphics{rao13b.ps}}}}
\caption{{\it Left, a:} Metallicity vs. redshift for DLAs
plotted as open squares where the area of each square is proportional
to $N(HI)$ (Rao et al. 2005b, Prochaska et al. 2003).  The cosmic mean
metallicities weighted by $N(HI)$, $\left<Z\right>$, are filled stars
and the unweighted mean metallicities are filled circles. The best fit
linear slope is $m=-0.26\pm0.06$, indicating 
increasing metallicity with time.  {\it Right, b:} Metal mass density
as a function of redshift from Hopkins et al. (2005). 
Solid circles are DLA metallicities from Figure 13a converted to
mass density, and the hatched region is derived from the SFH
corresponding to the hatched region in Figure 12. Triangles at
$z=0$ are from  Calura \& Matteucci (2004); open circles are
from Dunne et al. (2003). }
\end{figure}

However, the possibility that these very high column density gas
systems are being missed  by DLA surveys may not only be due to their
small gas cross sections, but also because they are likely to be  very
dusty. Ledoux et al. (2003) find that the DLAs in which H$_2$ is
detected  have among the highest metallicities and the highest
depletion factors, hinting at the possibility of much higher
depletions in much higher column density molecular  gas clouds. These
are not found in the radio loud quasar surveys for DLAs because these
surveys also suffer from the small cross section selection effect.
Not enough radio loud quasars have yet been surveyed to find the
putative one in a thousand very high column density system.

Gravitational lensing has the opposite effect on DLA surveys.
Magnification by DLA galaxies could brighten background quasars, and
preferentially include them in magnitude-limited samples. Le Brun et
al. (2000), with HST imaging observations, showed no evidence for
multiple images of background quasars and concluded that the quasars
were magnified by at most 0.3 magnitudes. In addition, Ellison et
al. (2004) and P\'eroux et al. (2004) using statistical tests on low
redshift MgII and DLA  samples, showed that lensing bias is a minor
effect. More recently, using the SDSS MgII survey results of Nestor
(2004), M\'enard et al. (2005) show that quasars behind strong MgII
absorbers, of which DLAs are a subset, show little magnification bias,
and that its effect on $\Omega_{DLA}$ at low redshift is
negligible (see also M\'enard 2005). 
It is also unlikely that the  lowest redshift points that
we derived from our HST-UV data (Figure 9) are affected by lensing
bias. This is because the DLAs with the highest HI column densities at
$z\approx 0.5$ arise in dwarf galaxies (Rao et al. 2003), and
consequently, do  not have the mass required to produce significant
magnification.

\section{Summary}

Since the discovery of DLAs in the early 1970s, a  considerable
amount of effort has been put into deciphering the  properties of
neutral gas in the universe. The observational techniques that have to
be used to trace the neutral gas separate the Ly$\alpha$ universe into
three redshift regimes: the optical/IR at $z>1.65$, the UV at
$0<z<1.65$, and  the radio (21 cm) at $z=0$.  The current status of
DLA surveys in these three redshift regimes was  reviewed; while the
$z=0$ and $z>1.65$ regimes have been more than adequately serviced by
nearly all-sky surveys, the UV regime could do with an order  of
magnitude more DLAs. We also put forward a change in the paradigm that
DLAs trace the bulk of the neutral gas in the universe. Selection
effects that preclude us from observing the highest column density gas
might be affecting our interpretation of what DLA statistics really
mean. Evidence  for this comes from a comparison of DLA and luminous
mass densities, SFRs, and metal mass densities as a
function of redshift. See Turnshek et al. (2005, these proceedings) for 
more discussion on DLA selection effects. 

\begin{acknowledgments}
I thank the organizing committee and  Shanghai Astronomical
Observatory for hosting a lively and productive conference. I would
like to acknowledge Dave Turnshek, Art Wolfe,  Chris Churchill, Andrew
Hopkins, Brice M\'enard, Jason Prochaska,  and  Martin Zwaan for helpful 
discussions.
This work was funded by grants from NASA-STScI, NASA-LTSA, and
NSF. HST-UV spectroscopy made the $N(HI)$ determinations possible.  We
thank members of the SDSS collaboration who made the SDSS project a
success. Funding for creation and distribution of the SDSS Archive has
been provided by the Alfred P. Sloan Foundation, Participating
Institutions, NASA, NSF, DOE, the Japanese Monbukagakusho, and the Max
Planck Society. The SDSS Web site is www.sdss.org. The SDSS is managed
by the Astrophysical Research Consortium for the Participating
Institutions: University of Chicago, Fermilab, Institute for Advanced
Study, the Japan Participation Group, Johns Hopkins University, Los
Alamos National Laboratory, the Max-Planck-Institute for Astronomy
(MPIA), the Max-Planck-Institute for Astrophysics (MPA), New Mexico
State University, University of Pittsburgh, Princeton University, the
United States Naval Observatory, and University of Washington.

\end{acknowledgments}

\begin{discussion}

\end{discussion}

\end{document}